\documentclass{amsart}

\usepackage{amsfonts,amssymb,amscd,amsmath,enumerate,verbatim,newlfont,calc}

\usepackage{amsfonts}

\let\origmaketitle\maketitle
\def\maketitle{
  \begingroup
  \let\MakeUppercase\relax 
  \origmaketitle
  \endgroup
}

\begin{document}

\title[Equations of motion for 2-body problem... (JDSGT 9(2), 2011)]{Equations of motion for two-body problem according to an observer inside the gravitational field \footnote { \\ \it Journal of Dynamical Systems $\&$ Geometric Theories \\Vol.9, Number 2 (2011) 115-135.
\\
\copyright Taru Publications}}

\author[K. Tren\v{c}evski and E. Celakoska]{ {\bf {Kostadin Tren\v{c}evski$^1$ and Emilija
Celakoska$^2$}}
\\{$^1$Faculty of Natural Sciences and Mathematics}
\\{Sts. Cyril and Methodius University, P.O.Box 162, Skopje, R. Macedonia}
\\ {$^2$Faculty of Mechanical Engineering}
\\{Sts. Cyril and Methodius University, P.O.Box 464, Skopje, R. Macedonia}
\\e-mails: kostatre@pmf.ukim.mk, emilija.celakoska@mf.edu.mk }

\maketitle

\begin{abstract}
In this paper we consider equations of motion for 2-body problem
according to an observer close to one of the gravitational bodies.
The influence of the Thomas precession of the observer's frame has
an important role. The equations of motion are based on a nonlinear
connection modified by the action of an orthogonal tensor which
synchronizes the 4-velocities of the considered two bodies. Finally
we present periastron shift according to an observer inside the
gravitational field, using the orthonormal coordinates. This is
different approach than that used in General Relativity where the
periastron shift is only given as observed far from the massive
bodies.
\end{abstract}


\

\noindent {\bf AMS Classification: 53B15, 53B50, 83D05, 70F25,
70F15}.
\\
{\bf Keywords: Thomas precession, nonlinear connection, gravitation,
2-body problem, nonholonomic coordinates} \vspace{1cm} 

\section{Introduction}

We present equations of motion based on a nonlinear connection
representing a relation between the gravitational source and test
particle. We use $ict$ convention (see pp. 51 in \cite{M} about
$ct/ict$ conventions). So, we work with the Euclidean metric
$\hbox{diag}(1,1,1,1)$ and upper and lower indices will not differ.
Indeed, we consider simply a Minkowski space, and shall use
orthonormal coordinates.

In the recent paper \cite{TC3} in flat Minkowski space it is
obtained a general formula for frequency redshift/blueshift based on
the 4-wave vector in flat Minkowski space, which simultaneously
explains the Doppler effect, gravitational redshift and under one
cosmological assumption it also explains the cosmological redshift
and the blueshift arising from the Pioneer anomaly. For this reason,
and the recent results about non-linear connection in Minkowski
space \cite{TC4}, it is not necessary to employ the apparatus of
curved spaces for gravitational research (like, for example, in
General Relativity), so, we also choose to use the nonlinear
connection in this paper. We will use only one gravitational
potential and denote it by $\mu =1+\frac{GM}{rc^2}$, instead of the
Newton gravitational potential $\frac{GM}{r}$.

We build the connection step by step, starting from simpler cases
and then, by including additional relevant variables, we achieve its
final form. The goal is to distinguish two general aspects:
equations of motion for an observer far from the massive bodies and
for an observer inside a gravitational field. The result is applied
to periastron shift of binaries. If the observer is far from the
massive bodies, our results are the same as in the General
Relativity (GR).

For the equations of motion it is also important the position of the
observer, i.e. whether he is away from the gravitational field, or
inside the gravitational field. So, we can distinguish four cases:

1. The observer is far from the massive bodies and the coordinates
are orthonormal;

2. The observer is inside the gravitational field and the
coordinates are ordinary;

3. The observer is inside the gravitational field and the
coordinates are orthonormal;

4. The observer is far from the massive bodies and the coordinates
are ordinary.

Here orthonormal coordinates means coordinates of flat Minkowski
space. In this paper, we focus mainly on the case 3. The first case
is considered in our another work \cite{TC4}. The case 4 (specially
the Einstein-Infeld-Hoffmann equations) is considered in the GR. In
this paper the similarities and differences among these four cases
will be emphasized.

\section{Some preliminaries about a non-linear connection}

Non-linear connection means that the property
$\nabla_{aX+bY}=a\nabla_X+b\nabla_Y$ is not satisfied. The reason is
simply because the 4-vectors of velocities do not form a vector
space and we will consider them as Lorentz boosts. The definition of
the 4-velocities is made in such a way that they correspond to
tangent vectors on geodesics on 4-dimensional manifold, but on the
other side, it is clear that the special-relativistic addition does
not support classical linear combinations. So, motions of objects in
a gravitational field would be more accurately described with
equations of motion using a nonlinear connection. The non-linear
connection was introduced in recent paper \cite{TC4} for observer
far from the massive bodies (case 1.) and now we shall extend it for
in order to study the case 3. We do that in three steps.

\subsection{Using an analogy from electromagnetism}

We will make a complete analogy with the electromagnetism, where
instead of the charge $e$ we will consider mass $M$, and instead of
the potential $\frac{e}{r}$ we will consider the gravitational
potential $\frac{GM}{r}$, assuming that $M$ has the same value in
each inertial coordinate system. Further we will introduce an
antisymmetric tensor $\phi_{ij}$ analogous to the tensor of
electromagnetic field. Let us consider the motion of a test body
with mass $m$ under a gravitational attraction of a body with mass
$M$. The 4-vector of velocity of the gravitational body is denoted
by
$$(U_{1},U_{2},U_{3},U_{4}) = \frac{1}{\sqrt {1-u^{2}/c^{2}}}\Bigl (
\frac{u_{x}}{ic}, \frac{u_{y}}{ic}, \frac{u_{z}}{ic}, 1\Bigr )
,\eqno{(2.1)}$$ where $\vec{u}=(u_{x},u_{y},u_{z})$ is the
corresponding 3-vector of velocity. The 4-vector of velocity of the
test body with mass $m$ is denoted by
$$ (V_{1},V_{2},V_{3},V_{4}) = \frac{1}{\sqrt {1-v^{2}/c^{2}}}\Bigl (
\frac{v_{x}}{ic}, \frac{v_{y}}{ic}, \frac{v_{z}}{ic}, 1\Bigr
).\eqno{(2.2)}$$

At every point $(x,y,z)$, for a stationary gravitational body with
point mass $M$, we define tensor $\phi$ with
$$(\phi_{ij})=\left[ \begin{array}{cccc}
0 & 0 & 0 & \frac{GM}{r^3c^2}(x-x_0)\\[1.9ex]
0 & 0 & 0 & \frac{GM}{r^3c^2}(y-y_0)\\[1.9ex]
0 & 0 & 0 & \frac{GM}{r^3c^2}(z-z_0)\\[1.9ex]
-\frac{GM}{r^3c^2}(x-x_0) & -\frac{GM}{r^3c^2}(y-y_0) &
-\frac{GM}{r^3c^2}(z-z_0) & 0\end{array}\right ] , \eqno{(2.3)}$$
where $(x_0,y_0,z_0)$ is the position of the gravitational body. So,
$c^2(\phi_{41},\phi_{42},\phi_{43})$ represents the Newton
acceleration toward the gravitational body. Considering its
placement in $\phi_{ij}$, it is analogous to the electric field
$\vec{E}$ in the electromagnetic tensor.

We can use Lorentz transformations for uniform motion of particles
and also, the principle of superposition of weak fields, and now
$\phi_{ij}$ is determined if all gravitational bodies are moving
uniformly.

For a gravitational body moving non-uniformly, we consider the
components of the tensor $\phi_{ij}$ at each space-time point as
analogous to the electromagnetic tensor derived by the
Lienard-Wiechert potentials. We assume that the gravitational
interaction transmits with velocity $c$. So, we get the following
analogous formulas as in electrodynamics
$$c^2(\phi_{41},\phi_{42},\phi_{43})=
-\frac{GM}{(R-\frac{\vec{R}\cdot \vec{u}}{c})^3} \Bigl
(\vec{R}-\frac{\vec{u}}{c}R\Bigr )- \frac{GM}{c^2
(R-\frac{\vec{R}\cdot \vec{u}}{c})^3}\vec{R}\times \Bigl [ \Bigl
(\vec{R}-\frac{\vec{u}}{c}R\Bigr )\times \dot{\vec{u}}\Bigr ],
\eqno{(2.4a)}$$
$$\frac{c}{i}(\phi_{32},\phi_{13},\phi_{21})=\frac{1}{R}\vec{R}\times
(\phi_{41},\phi_{42},\phi_{43}).\eqno{(2.4b)}$$ Here, $\vec{u}$ is
the velocity of the gravitational body, $\vec{R}$ is the 3-vector
from the gravitational body to the considered point $(x,y,z,ict)$ in
the chosen coordinate system calculated at the space-time point
$(x',y',z',ict')$ of the gravitational body, such that after time
$t-t'$ of transmission of the interaction, the effect arrives at the
considered point $(x,y,z,ict)$. Thus, $t'$ is solution of the
equation
$$t=t'+\frac{R(t')}{c}.\eqno{(2.5)}$$
In (2.4a) $\dot{\vec{u}}=\partial \vec{u}/\partial t'$ and $R=\vert
\vec{R}\vert$.

For uniform motion $\dot{\vec{u}}=0$, so the equation (2.4a) reduces
to
$$c^2(\phi_{41},\phi_{42},\phi_{43})=
-\frac{GM}{R^3}\vec{R}\frac{1-\frac{u^2}{c^2}} {\Bigl
(1-\frac{u^2}{c^2}\sin^2\theta \Bigr )^{3/2}},\eqno{(2.6)}$$ where
$\theta$ is the angle between $\vec{R}$ and $\vec{u}$, and $\vec{R}$
is the 3-vector from the gravitational body to the considered point
at time $t$.

Since for 2-body problem $\vec{R}$ is collinear with
$\dot{\vec{u}}$, the last term in (2.4a) can be neglected up to
$c^{-2}$. So we shall use only (2.6).

Although we accepted some facts from the electromagnetism, we must
emphasize that there are two essential differences, which will be
considered in the subsections 2.2 and 2.3.

i) While the charge $e$ in electrodynamics is invariant scalar in
all coordinate systems, here the mass $M$ is not invariant. The
reason is that the mass depends on the velocity of the body,
according to the Special Relativity. Thus, the tensor $\phi$ must be
modified.

ii) The Lorentz force acting on charged particles at the considered
point depends on the electromagnetic field and does not depend on
the velocity of the source of the electromagnetic field. Considering
gravitational forces, the motion depends on the velocity of the
source of gravitation.

\subsection{Role of the masses in gravitational force and in the tensor $\phi$}

A mass far from the massive bodies measured by an observer far from
gravitational influence will be called proper mass and will be denoted by $m$,
$M$, $m_1$, $m_2$,... An observer far from the massive bodies
observing a body with proper mass $m$ that has fallen into a
gravitational field with potential $\mu =1+\frac{GM}{Rc^2}$, will
measure the value $\frac{m}{1+\frac{GM}{Rc^2}}$ for the mass of the
body. For a test body with small mass $m$ with respect to the
gravitational body, the mass
$\frac{m}{1+\frac{GM}{Rc^2}}\frac{1}{\sqrt{1-\frac{v^2}{c^2}}}$ will
be unchanged up to $c^{-2}$ during motion. This is in accordance
with conservation of energy (potential + kinetic) in a gravitational
field.

Let us consider two bodies with masses $m_1$ and $m_2$ on a distance
$R$ between their centers. Then, the gravitational force originating
from the body with mass $m_2$ which acts on the body with mass
$m_1$, is assumed to be
$$\vec{f}=\frac{m_1}{1+\frac{Gm_2}{Rc^2}}\nabla
\frac{Gm_2}{R(1+\frac{G m_1}{Rc^2})},\eqno{(2.7)}$$ while the
acceleration of the body with mass $m_1$ is assumed to be
$$\vec{a}=\frac{1}{1+\frac{Gm_2}{Rc^2}}\nabla
\frac{Gm_2}{R(1+\frac{G m_1}{Rc^2})}.\eqno{(2.8)}$$ Up to $c^{-2}$,
the acceleration (2.8) can be written in the form
$$\vec{a}=-\frac{\vec{R}}{R}\frac{Gm_2}{R^2}
\Bigl (1-\frac{G(2m_1+m_2)}{Rc^2}\Bigr ).\eqno{(2.9)}$$

If we consider two bodies within a system of $n$-bodies, it is
necessary to use a more general formula for the acceleration. The
masses of two bodies with proper masses $m_1$ and $m_2$ in a
gravitational field are
$$\frac{m_1}{(1+\frac{Gm_2}{r_{12}c^2})(1+\frac{Gm_3}{r_{13}c^2})
(1+\frac{Gm_4}{r_{14}c^2})...},$$ and
$$\frac{m_2}{(1+\frac{Gm_1}{r_{12}c^2})(1+\frac{Gm_3}{r_{23}c^2})
(1+\frac{Gm_4}{r_{24}c^2})...}$$ respectively, where $r_{ij}$ is the
distance between the bodies with masses $m_i$ and $m_j$. Now,
analogously to (2.7) and (2.8), for the force/acceleration of the
body with mass $m_1$ we get
$$\vec{f}=\frac{m_1}{(1+\frac{Gm_2}{r_{12}c^2})(1+\frac{Gm_3}{r_{13}c^2})
(1+\frac{Gm_4}{r_{14}c^2})...}\nabla
\frac{Gm_2}{r_{12}(1+\frac{Gm_1}{r_{12}c^2})(1+\frac{Gm_3}{r_{23}c^2})
(1+\frac{Gm_4}{r_{24}c^2})...},\eqno{(2.10)}$$
$$\vec{a}=\frac{1}{(1+\frac{Gm_2}{r_{12}c^2})(1+\frac{Gm_3}{r_{13}c^2})
(1+\frac{Gm_4}{r_{14}c^2})...}\nabla
\frac{Gm_2}{r_{12}(1+\frac{Gm_1}{r_{12}c^2})(1+\frac{Gm_3}{r_{23}c^2})
(1+\frac{Gm_4}{r_{24}c^2})...}.\eqno{(2.11)}$$ Analogously to (2.9)
we obtain
$$\vec{a}=\Bigl [\Bigl (1+\frac{Gm_2}{r_{12}c^2}
\Bigr )\Bigl (1+\frac{Gm_1}{r_{12}c^2}\Bigr )^2 \Bigl
(1+\frac{Gm_3}{r_{13}c^2}\Bigr ) \Bigl
(1+\frac{Gm_3}{r_{23}c^2}\Bigr )\cdot$$

$$\cdot
\Bigl(1+\frac{Gm_4}{r_{14}c^2}\Bigr ) \Bigl
(1+\frac{Gm_4}{r_{24}c^2}\Bigr )...\Bigr ]^{-1} \nabla
\frac{Gm_2}{r_{12}}.\eqno{(2.12)}$$ The components of angular
velocity are much smaller than the components of acceleration in
$\phi$, so we can multiply the tensor $\phi$ by the coefficient
which stands in front of $\nabla \frac{Gm_2}{r_{12}}$ in (2.12).
This coefficient in (2.12) is a scalar in the Minkowskian space up
to $c^{-2}$, and hence the product will preserve the tensor
character of $\phi$.

For example, if $m_1$ is a negligible small mass and $m_2=M$ is
non-zero mass, then its acceleration is equal to
$$\vec{a}=-\frac{\vec{R}}{R}\frac{\frac{GM}{R^2}}
{1+\frac{GM}{Rc^2}}.\eqno{(2.13)}$$ This acceleration can be written
as
$$\vec{a}=c^2\nabla \ln \Bigl (1+\frac{GM}{Rc^2}\Bigr ).
\eqno{(2.14)}$$ Obviously, the potentials $\mu =1+\frac{GM}{Rc^2}$
and $C\mu $, where $C$ is a constant, lead to the same acceleration.

Now, let us consider the case when the observer is inside the
gravitational field, close to the body with mass $m_2$. We want to
determine the motion of the body with mass $m_1$ with respect to the
observer close to the body with mass $m_2$.
The gravitational force originating from the body with mass $m_2$
which acts on the body with mass $m_1$, now is given by
$$\vec{f}=\frac{m_1}{(1+\frac{Gm_2}{Rc^2})(1+\frac{Gm_1}{Rc^2})}\nabla
\frac{Gm_2}{R}\eqno{(2.15)}$$ instead of (2.7), while the
corresponding acceleration of the body with mass $m_1$ is given by
$$\vec{a}=\frac{1}{(1+\frac{Gm_2}{Rc^2})(1+\frac{Gm_1}{Rc^2})}\nabla
\frac{Gm_2}{R}.\eqno{(2.16)}$$ Namely, the observer sees that the
other body has mass
$\displaystyle\frac{m_1}{(1+\frac{Gm_2}{Rc^2})(1+\frac{Gm_1}{Rc^2})},$
because the potential
$$\displaystyle(1+\frac{Gm_2}{Rc^2})(1+\frac{Gm_1}{Rc^2})\approx1+\frac{G(m_1+m_2)}{Rc^2}$$
is such that its gradient yields the relative acceleration
$\displaystyle -\vec{R}\cdot\frac{G(m_1+m_2)}{R^2}$ of the body with
mass $m_1$ with respect to him. Therefore, for the mass in (2.15) we
used
$\displaystyle\frac{m_1}{(1+\frac{Gm_2}{Rc^2})(1+\frac{Gm_1}{Rc^2})}$.
The proper mass $m_2$ is observed to be unchanged with respect to
itself, and so in (2.15) the gradient is applied to
$\displaystyle\frac{Gm_2}{R}$. Now, analogously to (2.9), the
acceleration is
$$\vec{a}=-\frac{\vec{R}}{R}\frac{Gm_2}{R^2}
\Bigl (1-\frac{G(m_1+m_2)}{Rc^2}\Bigr ). \eqno{(2.17)}$$

\subsection{Influence of the velocity of the gravitational source
in the equations of motion}

Notice that in a system of four orthonormal vectors $A_{i1}$,
$A_{i2}$, $A_{i3}$ and $A_{i4}$, where $A_{i\alpha}$ is the $i$-th
coordinate of the $\alpha$-th vector, using that $A_{i\alpha}$ is an
orthogonal matrix, the following tensor
$$\frac{dA_{i\alpha}}{ds}A_{j\alpha}\eqno{(2.18)}$$
is also antisymmetric as $\phi_{ij}$ is defined to be. In (2.18)
$ds=ic\sqrt{1-\frac{v^2}{c^2}}dt$, just like in the Special
Relativity. For $U_i=V_i$, we identify $\phi_{ij}$ with the tensor
given in (2.18). Then, the physical interpretation of $\phi_{ij}$
can be obtained using the tensor (2.18).

So, $\phi$ is given by
$$ \phi = \left [\begin{array}{cccc}
0 & -i\omega _{z}/c & i\omega _{y}/c & -a_{x}/c^{2}\cr i\omega
_{z}/c & 0 & -i\omega _{x}/c & -a_{y}/c^{2}\cr -i\omega _{y}/c &
i\omega _{x}/c & 0 & -a_{z}/c^{2}\cr a_{x}/c^{2} & a_{y}/c^{2} &
a_{z}/c^{2} & 0\cr \end{array}\right ] , \eqno{(2.19)}$$ where
$\vec{a}=(a_{x},a_{y},a_{z})$ is the 3-vector of acceleration and
$\vec{w}= (w_{x},w_{y},w_{z})$ is the 3-vector of angular velocity.

We will introduce an orthogonal tensor $P(U,V)$, which should have
the properties: to be orthogonal and to transform the 4-velocity of
the source into the 4-velocity of the particle. It can be understood
as an apparatus for transition between the frames of the source and
the particle. The tensor $P=P(U,V)$ is given by
$$P_{ij}=\delta _{ij} - \frac{1}{1 + U_{s}V_{s}}
(V_{i}V_{j}+V_{i}U_{j}+U_{i}V_{j}+U_{i}U_{j}) +
2U_{j}V_{i}.\eqno{(2.20)}$$ We assume the equality
$$\frac{dA_{i\alpha}}{ds}A_{j\alpha}=
P_{ri}\phi_{rk}P_{kj},\eqno{(2.21)}$$ which would represent a
general formula for the parallel transport of the considered frame
$A_{i\alpha}$ in direction of the 4-vector of velocity $V_i$, or in
matrix form
$$\frac{dA}{ds}A^T=P^T\phi P.$$

The tensor $P_{ij}$ is an orthogonal matrix and it has the following
property $P(U,V)=P(V,U)^{-1}$. Some other properties of this tensor
and a justification for its appearance in (2.20) are given in
\cite{CTbug,TC}. For example, it is shown that using the standard
addition, one can not uniquely determine a 4-vector in the
Minkowskian space-time which would represent a relative 4-velocity
of a point $B$ with respect to a point $A$, assuming that $B$ moves
with 4-velocity $V$ and $A$ moves with 4-velocity $U$. So, the
tensor $P(U,V)$ provides a transition between frames, i.e.
$P_{ij}U_j=V_j$.

In the special case $(U_{i})=(0,0,0,1)$, the tensor $P(U,V)$ is
given by
$$P = \left [\begin{array}{cccc}
1-\frac{1}{\nu }V_{1}^{2} & -\frac{1}{\nu }V_{1}V_{2} &
-\frac{1}{\nu }V_{1}V_{3} & V_{1}\cr & & & \cr -\frac{1}{\nu
}V_{2}V_{1} & 1-\frac{1}{\nu }V_{2}^{2} & -\frac{1}{\nu }V_{2}V_{3}
& V_{2}\cr & & & \cr -\frac{1}{\nu }V_{3}V_{1} & -\frac{1}{\nu
}V_{3}V_{2} & 1-\frac{1}{\nu }V_{3}^{2} & V_{3}\cr & & & \cr -V_{1}
& -V_{2} & -V_{3} & V_{4}\cr \end{array}\right ], \eqno{(2.22)}$$
where $V_{1}, V_{2}, V_{3}, V_{4}$ are given by (2.2), $\nu
=1+V_{4}$, and this represents just a Lorentz transformation (as a
boost, without space rotation). If we multiply the equation (2.21)
by $A_{j\beta}$ and sum for $\beta =1,2,3,4$ we get
$$\frac{dA_{i\beta}}{ds}=P_{ri}\phi_{rk}P_{kj}A_{j\beta},$$
and hence for the parallel displacement of arbitrary (unit) vector
$A_i$ we get
$$\frac{dA_i}{ds}=P_{ri}\phi_{rk}P_{kj}A_j.\eqno{(2.23)}$$
Specially, for $A_i=V_i$, we obtain the equations for parallel
displacement, i.e.
$$\frac{dV_i}{ds}=P_{ri}\phi_{rk}P_{kj}V_j,\eqno{(2.24)}$$
and these are the equations of motion in orthonormal coordinates.

\section{Influence of the Thomas precession of the coordinate axes}

Let us return to the basic equations of motion (2.24). We mentioned
that these equations are Lorentz covariant, but although they give
the exact equations of motion, they are represented only {\em with
respect to the chosen coordinates}, which is a usual practice, but
it is not always the same with the observation with respect to the
distant stars.

Let us consider a gravitational body, a test body with zero mass and
a coordinate frame in a small neighborhood of the test body. In case
of weak gravitational field we may assume that the gravitational
field is caused by many small bodies (atoms) of the gravitational
body, which have zero angular momentums. Observing from the frame,
the distant stars would make an apparent (not true) rotation on the
sky with angular velocity
$$\vec{w}=\frac{1}{2}
\sum_i \frac{(\vec{v}-\vec{u}_i)\times \vec{a}_i}{c^2}\eqno{(3.1)}$$
where $\vec{u}_i$ is the velocity of the $i$-th body, $\vec{v}$ is
the velocity of the observer from the chosen coordinate system and
$\vec{a}_i$ is the Newtonian acceleration of the test body toward
the $i$-th gravitational body. This formula was obtained recently in
\cite{CJP} in order to obtain
Lorentz covariance of the precession of the axis of a gyroscope. %
Notice that (3.1) does not depend on the choice of the coordinate
system. This angular velocity of the distant stars on the sky is
observed by the telescope from the Gravity Prove B experiment, but
it is masked by much larger effects. Consequently, observed from a
system far from the massive bodies, which rests with respect to the
observer, i.e. $\vec{v}=0$, the coordinate frame rotates with the
opposite angular velocity
$$\vec{w}^*=\frac{1}{2}
\sum_i \frac{\vec{u}_i\times \vec{a}_i}{c^2}.\eqno{(3.2)}$$

Moreover, any coordinate system far from the massive bodies observes
that
the coordinate frame rotates with the same angular velocity (3.2) \cite{CJP}. %
What does it mean that one observer sees that another coordinate
system rotates with angular velocity $w$ \cite{CJP}? If we
have two observers from coordinate systems $S_1$ and $S_2$, and
assume that the precession of a gyroscope's spin axis is observed to
have angular velocity $w_1$ and $w_1-w_2$ from $S_1$ and $S_2$
respectively, then we may define that the observer from $S_1$ sees
that the "coordinate system $S_2$ rotates with angular velocity
$w_1-w_2$".

The angular velocity (3.2) is analogous to the Thomas precession.
While the Thomas precession is related with the gyroscopes, the
angular velocity (3.2) is related to the coordinate frame near the
massive bodies. We shall call this angular velocity also
Thomas precession. This effect will be also applied in case of
periastron shift in order to obtain the precession with respect to
the distant stars.
This is the main goal of this paper.%

How this precession of frames influence the equations of motion?
More precisely, what would be the modification/correction of the
equations of motion, so they would give equations of motion with
respect to a non-precessing coordinate system far from the massive
bodies? In the equations of motion (2.24), all possible influences
appearing in the chosen coordinate system are implicitly included.
This is a consequence of the covariance of these equations. The
angular velocity (3.2) is a real angular velocity, and its influence
should be subtracted from the tensor equation (2.24).

Now let us return to the required correction for the equations of
motion of any particle with velocity $\vec{v}$. The periastron shift
has to be Lorentz covariant, i.e. it should not depend on the
observer if he is in an inertial coordinate system far from the
massive bodies. For the considered particle we must also consider
the Coriolis acceleration and the transverse acceleration, while the
centrifugal acceleration which depends on $w^2$ will be of order
$c^{-4}$ and can be neglected. Notice that the considered
gravitational body should be considered as center of rotation. Thus
for example for the Coriolis force we should take into account the
relative velocity of the test particle with respect to the source of
gravitation. We find the acceleration of the considered body caused
by a particle with mass $M_i$ and then all such accelerations should
be summed. Hence, according to (3.2), the required correcting
acceleration is
$$\vec{a}_{cor}=\sum_i \Bigl (2(\vec{v}-\vec{u}_i)\times \vec{w}_i^*+\vec{R}_i\times
\frac{d\vec{w}_i^*}{dt}\Bigr ),$$ where
$\vec{w}^*_i=\frac{1}{2c^2}(\vec{u}_i\times
\frac{GM_i(-\vec{R}_i)}{R_i^3})$, $\vec{u}_i$ is the velocity of the
$i$-th body, $M_i$ is its mass, $\vec{R}_i$ is the vector from the
$i$-th body to the moving test body, $R_i=\vert \vec{R}_i\vert$ and
$\vec{v}$ is the velocity of the observed coordinate system.

After some transformations using double vector products, we obtain
$$\vec{a}_{cor}=\frac{G}{c^2}\sum_i \Bigl [\frac{M_i}{2R_i^3}
(\vec{v}-\vec{u}_i)(\vec{R}_i\cdot \vec{u}_i)+
\frac{M_i}{R_i^3}\vec{R}_i\Bigl ( \vec{u}_i\cdot
(\vec{v}-\vec{u}_i)-\frac{3}{2}\frac{((\vec{v}-\vec{u}_i)\cdot
\vec{R}_i) (\vec{u}_i\cdot \vec{R}_i)}{R_i^2}\Bigr )-$$
$$-\frac{M_i}{2R_i}\dot{\vec{u}}_i+
\frac{M_i}{2R_i^3}\vec{R}_i(\vec{R}_i\cdot \dot{\vec{u}}_i)\Bigr ].
\eqno{(3.3)}$$


\section{Periastron shift of binary systems}

A straightforward calculation of the matrix $S=P^T\phi P$, where
$\phi$ is given by (2.19) and $P$ is given by (2.22), leads to
$$S_{41}=-S_{14}=i{\frac{\omega _z}{c}}V_2-i{\frac{\omega _y}{c}}V_3+
{\frac{a_x}{c{}^2}} \left( V_4+{\frac{(V_1){}^2}{1+V_4} }\right)
+{\frac{a_y}{c{}^2}} {\frac{V_1V_2}{1+V_4}}+{\frac{a_z}{c{}^2}
}{\frac{V_1V_3}{1+V_4}},$$
$$S_{42}=-S_{24}=i{\frac{\omega _x}{c}}V_3-i{\frac{\omega _z}{c}}V_1+
{\frac{a_x}{c{}^2}} {\frac{V_1V_2}{1+V_4}} +{\frac{a_y}{c{}^2}
}\left( V_4+{\frac{(V_2){}^2}{1+V_4}}\right) +{\frac{a_z}{c{}^2} }
{\frac{V_2V_3}{1+V_4}},$$
$$S_{43}=-S_{34}=i{\frac{\omega _y}{c}}V_1-i{\frac{\omega _x}{c}}V_2+
{\frac{a_x}{c{}^2}} {\frac{V_1V_3}{1+V_4}}+{\frac{a_y}{c{}^2}
}{\frac{V_2V_3}{1+V_4}} +{\frac{a_z}{c{}^2}}\left(
V_4+{\frac{(V_3)^2}{1+V_4}}\right),$$
$$S_{32}=-S_{23}={\frac{a_z}{c{}^2}}V_2-{\frac{a_y}{c{}^2}}V_3+
i{\frac{\omega _x}{c}}\left( V_4+{\frac{(V_1){}^2}{1+V_4} }\right)
+i{\frac{\omega_y}{c}}{\frac{V_1V_2}{1+V_4}}+i{\frac{\omega_z}{c}}
{\frac{V_1V_3}{1+V_4}},$$
$$S_{13}=-S_{31}={\frac{a_x}{c{}^2}}V_3-{\frac{a_z}{c{}^2}}V_1+
i{\frac{\omega _x}{c}} {\frac{V_1V_2}{1+V_4}} +i{\frac{\omega
_y}{c}}\left( V_4+{\frac{(V_2){}^2}{1+V_4}}\right) +i{\frac{\omega
_z}{c}} {\frac{V_2V_3}{1+V_4}},$$
$$S_{21}=-S_{12}={\frac{a_y}{c{}^2}}V_1-{\frac{a_x}{c{}^2}}V_2+
i{\frac{\omega _x}{c}} {\frac{V_1V_3}{1+V_4}}+i{\frac{\omega
_y}{c}}{\frac{V_2V_3}{1+V_4}} +i{\frac{\omega _z}{c}}\left(
V_4+{\frac{(V_3){}^2}{1+V_4}}\right) ,$$
$$S_{11}=S_{22}=S_{33}=S_{44}=0.\eqno{(4.1)}$$

We will consider the periastron shift for a pulsar and its companion
and denote by $m$ the mass of a pulsar and by $M$ the mass of its
companion, assuming that both bodies are moving in the same plane
($xy$-plane). We choose a coordinate system such that its origin
always passes through the line connecting the two bodies. Let
$(x,y,0)$ be the coordinates of the pulsar, and let $(x',y',0)$ be
the coordinates of its companion. We denote the 4-vector of velocity
of the pulsar by
$$(V_i)=\frac{1}{\sqrt{1-\frac{v^2}{c^2}}}
\Bigl (\frac{v_x}{ic},\frac{v_y}{ic},\frac{v_z}{ic},1\Bigr ),
\eqno{(4.2)}$$ and the 4-vector of velocity of its companion by
$$(U_i)=\frac{1}{\sqrt{1-\frac{u^2}{c^2}}}
\Bigl (\frac{u_x}{ic},\frac{u_y}{ic},\frac{u_z}{ic},1\Bigr ) .
\eqno{(4.3)}$$ It is convenient to use the notation
$R=\sqrt{(x-x')^2+(y-y')^2}$, $r=\sqrt{x^2+y^2}$, $\rho =1/r$ and
$r\approx \frac{M}{M+m}R$. If we make the replacements
$\displaystyle\frac{x-x'}{R}=\cos \alpha $ and
$\displaystyle\frac{y-y'}{R}=\sin \alpha$, then $\cos \alpha =
\displaystyle\frac{x}{r}$, $\sin \alpha = \displaystyle\frac{y}{r}$,
$x'/y'=x/y$, $u_x\approx\displaystyle -v_x\frac{m}{M}$, and
$u_y\approx\displaystyle -v_y\frac{m}{M}$.

The basic equations of motion in orthonormal coordinates are
$$\frac{dV_i}{ds}=S_{ij}V_j,$$
where $V_i$ is given by (4.2) and $ds=ic\sqrt{1-\frac{v^2}{c^2}}.$
These equations can be simplified into the following form
$$\frac{d^2x}{dt^2}=-\sqrt{1-\frac{v^2}{c^2}}(S_{41}v_x^2+S_{42}v_xv_y+
S_{43}v_xv_z)+ic\sqrt{1-\frac{v^2}{c^2}}(S_{12}v_y+S_{13}v_z+icS_{14}),
\eqno{(4.4a)}$$
$$\frac{d^2y}{dt^2}=-\sqrt{1-\frac{v^2}{c^2}}(S_{41}v_xv_y+S_{42}v_y^2+
S_{43}v_yv_z)+ic\sqrt{1-\frac{v^2}{c^2}}(S_{21}v_x+S_{23}v_z+icS_{24}),
\eqno{(4.4b)}$$
$$\frac{d^2z}{dt^2}=-\sqrt{1-\frac{v^2}{c^2}}(S_{41}v_xv_z+S_{42}v_yv_z+
S_{43}v_z^2)+ic\sqrt{1-\frac{v^2}{c^2}}(S_{31}v_x+S_{32}v_y+icS_{34}),
\eqno{(4.4c)}$$
$$\frac{d}{dt}\frac{1}{\sqrt{1-\frac{v^2}{c^2}}}=
S_{41}v_x+S_{42}v_y+S_{43}v_z,\eqno{(4.4d)}$$ where
$d^2x/dt^2=dv_x/dt$, $d^2y/dt^2=dv_y/dt$ and $d^2z/dt^2=dv_z/dt$.
Specially for the 2-body problem in the $xy$-plane we can replace
$S_{34}=S_{13}=S_{23}=0$, $v_z=0$ and we obtain the following simple
system
$$\frac{d^2x}{dt^2}=-\sqrt{1-\frac{v^2}{c^2}}(S_{41}v_x^2+S_{42}v_xv_y)
+ic\sqrt{1-\frac{v^2}{c^2}}S_{12}v_y-S_{14}c^2\sqrt{1-\frac{v^2}{c^2}},$$
$$\frac{d^2y}{dt^2}=-\sqrt{1-\frac{v^2}{c^2}}(S_{41}v_xv_y+S_{42}v_y^2)
-ic\sqrt{1-\frac{v^2}{c^2}}S_{12}v_x-S_{24}c^2\sqrt{1-\frac{v^2}{c^2}}.$$
Up to $c^{-2}$, we get
$$\frac{d^2x}{dt^2}=-\Bigl (1-\frac{v^2}{2c^2}\Bigr )c^2S_{14}-
\frac{v_x}{c^2}(a_xv_x+a_yv_y)+icv_yS_{12},\eqno{(4.5a)}$$
$$\frac{d^2y}{dt^2}=-\Bigl (1-\frac{v^2}{2c^2}\Bigr )c^2S_{24}-
\frac{v_y}{c^2}(a_xv_x+a_yv_y)-icv_xS_{12}.\eqno{(4.5b)}$$

Further, the components of the matrix $S=P(U,V)^T\phi P(U,V)$ should
be calculated analogously as in (4.1). In order to avoid large
expressions with accuracy up to $c^{-2}$, it is sufficient to use
the components (4.1), replacing $v_x$ by $v_x-u_x$ and $v_y$ by
$v_y-u_y$. Hence, for $S_{14}$, $S_{24}$, and $S_{12}$ we obtain
$$S_{14}=\Bigl [-a_x\Bigl (\frac{1}{\sqrt{1-\frac{(\vec{v}-\vec{u})^2}{c^2}}}-
\frac{(v_x-u_x)^2}{2c^2}\Bigr )+a_y\frac{(v_x-u_x)(v_y-u_y)}{2c^2}-w_z(v_y-u_y)\Bigr
]\frac{1}{c^2},$$
$$S_{24}=\Bigl [-a_y\Bigl (\frac{1}{\sqrt{1-\frac{(\vec{v}-\vec{u})^2}{c^2}}}-
\frac{(v_y-u_y)^2}{2c^2}\Bigr )+a_x\frac{(v_x-u_x)(v_y-u_y)}{2c^2}+w_z(v_x-u_x)\Bigr
]\frac{1}{c^2},$$
$$S_{12}=-\frac{i}{c}\Bigl [\frac{a_x}{c^2}(v_y-u_y)-
\frac{a_y}{c^2}(v_x-u_x)+w_z\Bigr ] .$$

According to (2.19), (2.6), (2.17), up to $c^{-2}$,
$$\displaystyle \frac{1-\frac{u^2}{c^2}}{(1-\frac{u^2}{c^2}\sin^2\theta)^{3/2}}
=\frac{1-\frac{u^2}{c^2}}
{(1-\frac{u^2}{c^2})^{3/2}(1+\frac{u^2}{c^2}\cos^2\theta)^{3/2}}=$$
$$=\frac{1}{\sqrt{1-\frac{u^2}{c^2}}}\frac{1}{(1+\frac{1}{c^2}(\vec{u}\cdot
\frac{\vec{r}}{r})^2)^{3/2}}= \displaystyle
\frac{1}{\sqrt{1-\frac{u^2}{c^2}}}\frac{1}{(1+\frac{1}{c^2}\frac{m^2}{M^2}
(\vec{r}'\cdot \frac{\vec{r}}{r})^2)^{3/2}}=$$
$$=\frac{1}{\sqrt{1-\frac{u^2}{c^2}}}\frac{1}{(1+\frac{1}{c^2}\frac{m^2}{M^2}
(\frac{dr}{dt})^2)^{3/2}}= \displaystyle
\frac{1}{\sqrt{1-\frac{u^2}{c^2}}}\frac{1}{(1+\frac{1}{c^2}\frac{m^2}{(M+m)^2}
(\frac{dR}{dt})^2)^{3/2}},$$ the components $a_x$, $a_y$, and $w_z$
are
$$a_x=-\frac{x}{r}\frac{1}{\sqrt{1-\frac{u^2}{c^2}}}\frac{GM}{R^2}
\Bigl (1-\frac{G(M+m)}{Rc^2}\Bigr )\lambda^{-3},$$
$$a_y=-\frac{y}{r}\frac{1}{\sqrt{1-\frac{u^2}{c^2}}}\frac{GM}{R^2}
\Bigl (1-\frac{G(M+m)}{Rc^2}\Bigr )\lambda^{-3},$$
$$w_z=\frac{Gm}{R^2c^2}\frac{v_xy-v_yx}{r}\lambda^{-3},$$
where $\lambda =\sqrt{1+\displaystyle\frac{1}{c^2}
\frac{m^2}{(M+m)^2}\Bigl (\frac{dR}{dt}\Bigr )^2}$. Further, we will
omit this coefficient in expressions of type $c^{-2}$. Now, having
the system of equations (4.5a) and (4.5b) for the motion of a body
with mass $m$ influenced by the gravitation of a body with mass $M$,
we can calculate the periastron shift in two steps. The first step
is consisted in finding an equation analogous to perihelion shift in
orthonormal coordinates. In the second step we sum the equation
(4.5a) multiplied by $2v_x$ and the equation (4.5b) multiplied by
$2v_y$. That equation can be integrated and the value of $v^2$ can
be found. After these two steps the periastron shift can be
obtained. We present only the final results of these two steps
avoiding the long algebraic and differential calculations.

We distinguish three aspects for obtaining the periastron shift and
we number them by i), ii) and iii).
In i) and ii) will be calculated the periastron shift with respect
to the chosen coordinate system, using the non-linear connection
from section 2. In iii) will be calculated periastron shift using
the results from section 3.

i) We use the orthonormal coordinate system and we {\em assume a
priori that the two bodies and the coordinate center are collinear}.
In this case we can continue to deal with the system (4.5a,b) as
previously described. Using the equalities between $v_x$, $u_x$;
$v_y$, $u_y$; $r$, $R$ and so on, the system (4.5) can be reduced to
the following form
$$\frac{d^2\vec{r}}{dt^2}=-\frac{\vec{R}}{R}\frac{GM}{R^2}\Bigl [
1+\frac{V^2}{c^2}\frac{M^2+4Mm+2m^2}{(M+m)^2}-\frac{G(M+m)}{Rc^2}-$$
$$-\frac{3}{2c^2} \frac{m^2}{(M+m)^2}\Bigl (\frac{dR}{dt}\Bigr
)^2\Bigr ]+\vec{V}\frac{dR}{dt}\frac{GM}{R^2} \Bigl
(\frac{M}{M+m}+\frac{3}{2}\Bigr )\frac{1}{c^2},\eqno{(4.6)}$$ where
$\vec{V}$ is the relative velocity of the pulsar with respect to its
companion.

The first step ahead from the system (4.6) yields the following
equation
$$\Bigl (\frac{d\rho}{d\varphi}\Bigr )^2+\rho^2=v^2C_2^{-2}
\Bigl [1+\frac{5M+3m}{M+m}\frac{GM\rho}{c^2}\Bigr ].\eqno{(4.7)}$$
The second step is more complicated. Although $\lambda$ has a
significant role in $v^2$, $\lambda $ has no influence in (4.7) and
it has no role in the periastron shift \cite{TC4}. So, we will omit
$\frac{3}{2c^2}\frac{m^2}{(M+m)^2}\Bigl (\frac{dR}{dt}\Bigr )^2$ in
(4.6), i.e. we consider $\lambda =1$. The second step yields
$$v^2=2\frac{GM^3\rho}{(M+m)^2}-4\frac{M^6G^2\rho^2}{(M+m)^4c^2}
-2\frac{M^5mG^2\rho^2}{(M+m)^4c^2}+
\frac{C\rho}{c^2}+K,\eqno{(4.8)}$$ where $C$ and $K$ are two
mutually dependent constants, which do not play any role in the
periastron shift. We get
$$\Bigl (\frac{d\rho}{d\varphi}\Bigr )^2+\rho^2=A+B\rho +
\frac{6G^2M^4}{c^2C_2^2(M+m)^2} \cdot \Bigl
(1+\frac{1}{3}\frac{Mm}{(M+m)^2}\Bigr )\rho^2.$$

Since $C_2=\frac{2}{P}\pi a^2\sqrt{1-\epsilon^2}=\frac{2\pi}{P}
a_r^2\frac{M^2}{(M+m)^2}\sqrt{1-\epsilon^2}$, where $a_r$ is the
semi-major axis of the relative orbit and $P$ is the orbital period,
we obtain
$$\Bigl (\frac{d\rho}{d\varphi}\Bigr )^2+\rho^2=A+B\rho +
\frac{3P^2G^2}{2\pi^2c^2}\frac{(M+m)^2}{a_r^4(1-\epsilon^2)} \cdot
\Bigl (1+\frac{1}{3}\frac{Mm}{(M+m)^2}\Bigr )\rho^2.$$ So, for the
periastron shift we obtain
$$\Delta \varphi = \frac{3G^2(M+m)^2P^2}{2\pi c^2a_r^4(1-\epsilon^2)}
\cdot \Bigl (1+\frac{1}{3}\frac{Mm}{(M+m)^2}\Bigr ).$$ Using that
$a_r^3=P^2G(M+m)/(4\pi ^2)$, we obtain finally
$$\Delta \varphi = \frac{6\pi (4\pi^2)^{1/3}G^{2/3}}
{P^{2/3}c^2(1-\epsilon^2)}(M+m)^{2/3} \cdot \Bigl
(1+\frac{1}{3}\frac{Mm}{(M+m)^2}\Bigr ).\eqno{(4.9)}$$

ii) The equations of motion of the body with mass $m$ are given by
(4.6). Analogously to these equations, the equations of motion of
the body with mass $M$ (pulsar companion) are given by
$$\frac{d^2\vec{r}'}{dt^2}=\frac{\vec{R}}{R}\frac{Gm}{R^2}\Bigl [
1+\frac{V^2}{c^2}\frac{m^2+4Mm+2M^2}{(M+m)^2}-\frac{G(M+m)}{Rc^2}-$$
$$-\frac{3}{2c^2} \frac{M^2}{(M+m)^2}\Bigl (\frac{dR}{dt}\Bigr
)^2\Bigr ]-\vec{V}\frac{dR}{dt}\frac{Gm}{R^2} \Bigl
(\frac{m}{M+m}+\frac{3}{2}\Bigr )\frac{1}{c^2}.\eqno{(4.10)}$$
Subtracting the equation (4.10) from (4.6), after some
transformations we get
$$\frac{d^2\vec{R}}{dt^2}=-\frac{\vec{R}}{R}\frac{G(M+m)}{R^2}\Bigl [
1+\frac{V^2}{c^2}\frac{M^2+5Mm+m^2}{(M+m)^2}-\frac{G(M+m)}{Rc^2}-$$
$$-\frac{3}{2c^2} \frac{Mm}{(M+m)^2}\Bigl (\frac{dR}{dt}\Bigr
)^2\Bigr
]+\vec{V}\frac{dR}{dt}\frac{G}{R^2}\frac{5M^2+6Mm+5m^2}{2(M+m)c^2}.\eqno{(4.11)}$$
While the assumption that the two bodies and the coordinate center
are collinear was essential for deriving the equations (4.6) and
(4.10), it has no role in the system (4.11), but (4.11) has a
weakness because it is a subtraction between (4.6) and (4.10) which are
deduced for different observers (coordinate systems). The equation
(4.11) is independent of the coordinate system, it depends only on
the relative parameters of the system.

A step ahead from the system (4.11), analogously to (4.7), yields
the following equation
$$\Bigl (\frac{d\frac{1}{R}}{d\varphi}\Bigr )^2+\frac{1}{R^2}=V^2C_2^{-2}
\Bigl [1+\frac{5M^2+6Mm+5m^2}{M+m}\frac{G}{Rc^2}\Bigr
].\eqno{(4.12)}$$

Analogously to i) we can ignore the term
$\frac{3}{2c^2}\frac{Mm}{(M+m)^2}\Bigl (\frac{dR}{dt}\Bigr )^2$ in
(4.11), because it has no role in the periastron shift and we obtain
$$V^2=2\frac{G(M+m)}{R}-\frac{G^2}{R^2c^2}(4M^2-2mM+4m^2)
+ \frac{C}{Rc^2}+K,\eqno{(4.13)}$$ where $C$ and $K$ are mutually
dependent constants which have no role in the periastron shift.
Hence, we come again to the same formula (4.9).

iii) Now, we will calculate the periastron shift in Lorentz
invariant form where the Thomas precession from section 3 will be
used. The results from section 3 must be used because (4.9) was
deduced by assuming a priori that the two bodies and the coordinate
origin are collinear. This includes an additional precession of the
axis between the pulsar and the companion which should be subtracted
from (4.9). The precession (4.9) should be corrected in the
following way. Let $O$ be an observer who rests with respect to the
baricenter of the two bodies from an inertial coordinate system far
from the massive bodies. According to (3.2), $O$ observes the
coordinate frame close to the pulsar that rotates with angular
velocity
$$\frac{1}{2}\vec{u}\times \frac{-\vec{R}}{R^3}\frac{GM}{c^2},$$
while the coordinate frame close to the companion star which rotates,
with angular velocity
$$\frac{1}{2}\vec{v}\times \frac{\vec{R}}{R^3}\frac{Gm}{c^2}.$$
Each of these two angular velocities can be written via the
parameters of the relative orbit, i.e. each of them is equal to
$$\frac{1}{2}\vec{V}\times \frac{\vec{R}}{R^3}\frac{G(M+m)}{c^2}\cdot
\frac{Mm}{(M+m)^2},\eqno{(4.14)}$$ where $\vec{V}$ is the relative
velocity of the pulsar with respect to its companion. So the line
which connects the pulsar and its companion is observed to rotate
additionally with angular velocity which is opposite of the sum of
these two angular velocities, i.e. with
$$-\vec{V}\times \frac{\vec{R}}{R^3}\frac{G(M+m)}{c^2}\cdot
\frac{Mm}{(M+m)^2}.$$ These two angular velocities do not change the
distance from the pulsar to its companion, so they participate in
the periastron shift observed via the coordinate system. The sum of
both angular velocities should be integrated for a time of one
orbital period. After some standard calculations, this yields the
angle
$$\frac{1}{3}\frac{mM}{(M+m)^2}
\frac{3G^2(M+m)^2P^2}{2\pi c^2a_r^4(1-\epsilon^2)}=
\frac{1}{3}\frac{mM}{(M+m)^2} \frac{6\pi (4\pi^2)^{1/3}G^{2/3}}
{P^{2/3}c^2(1-\epsilon^2)}(M+m)^{2/3}$$ per orbit. This angle is
included in the total periastron shift (4.9) obtained via the
covariant equations of motion, and so after its subtraction from
(4.9) we obtain the precession
$$\Delta \varphi = \frac{6\pi (4\pi^2)^{1/3}G^{2/3}}
{P^{2/3}c^2(1-\epsilon^2)}(M+m)^{2/3},\eqno{(4.15)}$$ according to
observer $O$. This formula (4.15) for the periastron shift is the
same which predicts the GR, and it depends on the sum of the masses
$M+m$.

If the system of two bodies moves with a constant velocity
$\vec{v}_0$, then both velocities $\vec{v}$ and $\vec{u}$ should be
replaced by $\vec{v}+\vec{v}_0$ and $\vec{u}+\vec{v}_0$. In this
case the moving observer observes the angular velocity
$$-\frac{1}{2}(\vec{u}+\vec{v}_0)\times \frac{-\vec{R}}{R^3}\frac{GM}{c^2}
-\frac{1}{2}(\vec{v}+\vec{v}_0)\times
\frac{\vec{R}}{R^3}\frac{Gm}{c^2}=$$
$$=-\vec{V}\times \frac{\vec{R}}{R^3}\frac{G(M+m)}{c^2}\cdot
\frac{Mm}{(M+m)^2}+ \frac{1}{2}\vec{v}_0\times \Bigl
(\frac{\vec{R}}{R^3}\frac{GM}{c^2}-
\frac{\vec{R}}{R^3}\frac{Gm}{c^2}\Bigr ).$$ Thus, an additional
anomalous angular velocity
$$\frac{1}{2}\vec{v}_0\times \Bigl (\frac{\vec{R}}{R^3}\frac{GM}{c^2}-
\frac{\vec{R}}{R^3}\frac{Gm}{c^2}\Bigr )\eqno{(4.16)}$$ is observed.
So the main problem is to consider the perturbations which come from
(4.16).

First, notice that if $M=m$, then this angular velocity vanishes. In
case of pulsar and its companion star, very often it is $M\approx m$
and hence (4.16) is almost 0.

In general case, the constant vector $\vec{v}_0$ can be decomposed
as a sum of two vectors: a component which lies in the plane of
rotation of the binary system, and component which is orthogonal to
that plane. We will consider these two special cases separately.

Assume that the vector $\vec{v}_0$ lies in the orbital plane.
Standard calculation shows that the integral of the vector (4.16)
for a time of one period is zero. So, the total periastron shift
remains unchanged with respect to the moving observer, i.e. it is
given by (4.15) also.

Assume that the vector $\vec{v}_0$ is orthogonal to the orbital
plane. Then we have angular velocities in different planes, which
are orthogonal to the orbital plane. We shall see now that the
angular velocity (4.16) is a consequence from the Special
Relativity. Let us denote by $P$, $C$, $B$, and $O$ the pulsar, its
companion, the barycenter and the moving observer respectively. For
the sake of simplicity we assume that the eccentricity of the orbit
is 0, i.e. the velocities of the pulsar and its companion are
orthogonal to their radius-vectors. We denote the line $OB$ as
$z$-axis, and denote by $\Sigma$ the plane thought $B$ which is
orthogonal to $OB$. Now let us consider a composition of two Lorentz
transformations: the first motion in the $z$-axis with velocity
$v_0$ (motion of the barycenter with respect to the observer), and
motion with velocity $v$ in a direction orthogonal to the $z$-axis
(motion of the pulsar with respect to the barycenter). The composite
Lorentz transformation shows that according to the observer $O$, the
angle $\angle OBP$ is not right, but there is a departure of
$\frac{v_0v}{2c^2}$. Hence the pulsar moves in a plane which is on
distance $rvv_0/(2c^2)$ from the plane $\Sigma$. Analogously, the
companion star moves on a parallel plane which is on distance
$(R-r)uv_0/(2c^2)$ from the plane $\Sigma$. Notice that here
$v=\vert \vec{v}\vert$ and $u=\vert \vec{u}\vert$. Thus the angle
between the axis $CP$ and $\Sigma$ is equal to
$$\frac{1}{R}\Bigl (\frac{rvv_0}{2c^2}-\frac{(R-r)uv_0}{2c^2}\Bigr )=
\frac{v_0}{2c^2}\Bigl (\frac{r}{R}v-\frac{R-r}{R}u\Bigr )=$$
$$=\frac{v_0}{2c^2}\Bigl
(\frac{M^2}{(M+m)^2}V-\frac{m^2}{(M+m)^2}V\Bigr )=
\frac{1}{2}\frac{M-m}{M+m}\frac{v_0V}{c^2},$$ where $V$ is the
relative velocity of the pulsar with respect to its companion. The
last equality in vector form can be written as
$$-\frac{1}{2}\frac{M-m}{M+m}\frac{\vec{v}_0\times \vec{V}}{c^2}.$$
Now if we differentiate this term by $t$ we obtain (4.16).

Hence the conclusion that $O$ observes the same trajectories of $P$
and $C$ but in different planes, and hence the periastron shift
remains the same as (4.15) with respect to the observer $O$. The
moving observer sees that the orbital plane is not a fixed plane in
the space. This consideration confirms the statement that {\it the
"perturbations" arising from (4.16) and (3.2) have an important role
to "cover" the equations up to Lorentz invariant equations of motion
(2.24)}.

Now we will obtain the same conclusion using the formula for
acceleration (3.3). Using the equality $\vec{v}:\vec{u}=-M:m$, the
corrected acceleration for the pulsar is
$$\vec{a}_P= \frac{G(M+m)}{2c^2R^3}
\vec{v}(\vec{R}\cdot \vec{u})+\frac{G(M+m)}{c^2R^3}\vec{R}\Bigl (
\vec{u}\cdot \vec{v}-\frac{3}{2}\frac{(\vec{v}\cdot \vec{R})
(\vec{u}\cdot \vec{R})}{R^2}\Bigr ).\eqno{(4.17)}$$ Symmetrically,
for the corrected acceleration of the companion star we get
$$\vec{a}_C= \frac{G(M+m)}{2c^2R^3}
\vec{u}((-\vec{R})\cdot
\vec{v})+\frac{G(M+m)}{c^2R^3}(-\vec{R})\Bigl ( \vec{u}\cdot
\vec{v}-\frac{3}{2}\frac{(\vec{v}\cdot (-\vec{R})) (\vec{u}\cdot
(-\vec{R}))}{R^2}\Bigr ),$$ i.e.
$$\vec{a}_C=-\vec{a}_P.\eqno{(4.18)}$$
After these corrections of the accelerations of both bodies, the
lines which pass through them will not intersect at the coordinate
origin as previously. Thus, we can consider the periastron shift
only via the relative orbit. The correction of the relative
acceleration $\vec{a}_R=\vec{a}_P-\vec{a}_C$ can be written in the
form
$$\vec{a}_R=\Bigl \{\frac{G(M+m)}{c^2R^3}
\vec{v}(\vec{R}\cdot \vec{u})+\frac{G(M+m)}{c^2R^3}\vec{R}\Bigl (
\vec{u}\cdot \vec{v}-3\frac{(\vec{v}\cdot \vec{R}) (\vec{u}\cdot
\vec{R})}{R^2}\Bigr )\Bigr \}+\frac{G(M+m)}{c^2R^3}\vec{R}(\vec{u}\cdot \vec{v}).$$ The first
component
$$\frac{G(M+m)}{c^2R^3}
\vec{v}(\vec{R}\cdot \vec{u})+\frac{G(M+m)}{c^2R^3}\vec{R}\Bigl (
\vec{u}\cdot \vec{v}-3\frac{(\vec{v}\cdot \vec{R}) (\vec{u}\cdot
\vec{R})}{R^2}\Bigr )$$ does not change the periastron shift, while
the second component
$$\frac{G(M+m)}{c^2R^3}\vec{R}(\vec{u}\cdot \vec{v})\eqno{(4.19)}$$
leads to the angle
$$\frac{1}{3}\frac{mM}{(M+m)^2}
\frac{3G^2(M+m)^2P^2}{2\pi c^2a_r^4(1-\epsilon^2)}=
\frac{1}{3}\frac{mM}{(M+m)^2} \frac{6\pi (4\pi^2)^{1/3}G^{2/3}}
{P^{2/3}c^2(1-\epsilon^2)}(M+m)^{2/3}$$ per orbit. This angle is
included in the total periastron shift (4.9), and after its
subtraction from (4.9) we obtain (4.15).

{\bf Remark.} If the observer is far from the massive bodies, the
coordinate center coincides with the barycenter of the two bodies,
while the periastron shift between the barycenter (coordinate
origin) \cite{TC4} and each of the two bodies is not the same with
the periastron shift for the two bodies. In this paper, if the
observer is one of the bodies, ignoring the correction arising from
the Thomas precession we have the opposite situation: the barycenter
is not a fixed point and does not coincide with the coordinate
origin at each moment, but the periastron shift between the
coordinate origin and each of the bodies is the same as the
periastron shift between the two bodies (cases i) and ii)).

\end{document}